\documentclass[twocolumn,prl,reprint,superscriptaddress]{revtex4}
\usepackage{amsmath,amssymb}
\usepackage{graphicx}
\usepackage{float}
\usepackage{subfigure}
\usepackage{verbatim}

\usepackage{amsfonts}
\usepackage{dcolumn}
\usepackage{bm}
\usepackage{array,color}
\usepackage[colorlinks,citecolor=blue]{hyperref}
\usepackage[english]{babel}
\usepackage{braket}
\allowdisplaybreaks[4]
\begin{document}
\hyphenpenalty=5000
\tolerance=1000

\title{
Optical demonstration of quantum fault-tolerant threshold
}

\author{Kai Sun}
\affiliation{CAS Key Laboratory of Quantum Information, University of Science and Technology of China, Hefei 230026, People's Republic of China}
\affiliation{CAS Center For Excellence in Quantum Information and Quantum Physics, University of Science and Technology of China, Hefei 230026, People's Republic of China}

\author{Jin-Shi Xu}
\email{jsxu@ustc.edu.cn}
\affiliation{CAS Key Laboratory of Quantum Information, University of Science and Technology of China, Hefei 230026, People's Republic of China}
\affiliation{CAS Center For Excellence in Quantum Information and Quantum Physics, University of Science and Technology of China, Hefei 230026, People's Republic of China}

\author{Xiao-Ye Xu}
\affiliation{CAS Key Laboratory of Quantum Information, University of Science and Technology of China, Hefei 230026, People's Republic of China}
\affiliation{CAS Center For Excellence in Quantum Information and Quantum Physics, University of Science and Technology of China, Hefei 230026, People's Republic of China}

\author{Yong-Jian Han}
\email{smhan@ustc.edu.cn}
\affiliation{CAS Key Laboratory of Quantum Information, University of Science and Technology of China, Hefei 230026, People's Republic of China}
\affiliation{CAS Center For Excellence in Quantum Information and Quantum Physics, University of Science and Technology of China, Hefei 230026, People's Republic of China}

\author{Chuan-Feng Li}
\email{cfli@ustc.edu.cn}
\affiliation{CAS Key Laboratory of Quantum Information, University of Science and Technology of China, Hefei 230026, People's Republic of China}
\affiliation{CAS Center For Excellence in Quantum Information and Quantum Physics, University of Science and Technology of China, Hefei 230026, People's Republic of China}

\author{Guang-Can Guo}
\affiliation{CAS Key Laboratory of Quantum Information, University of Science and Technology of China, Hefei 230026, People's Republic of China}
\affiliation{CAS Center For Excellence in Quantum Information and Quantum Physics, University of Science and Technology of China, Hefei 230026, People's Republic of China}

\begin{abstract}
A major challenge in practical quantum computation is the ineludible errors caused by the interaction of quantum systems with their environment. Fault-tolerant schemes, in which logical qubits are encoded by several physical qubits, enable correct output of logical qubits under the presence of errors. However, strict requirements to encode qubits and operators render the implementation of a full fault-tolerant computation challenging even for the achievable noisy intermediate-scale quantum technology. Here, we experimentally demonstrate the existence of the threshold in a special fault-tolerant protocol. Four physical qubits are implemented using 16 optical spatial modes, in which 8 modes are used to encode two logical qubits. The experimental results clearly show that the probability of correct output in the circuit, formed with fault-tolerant gates, is higher than that in the corresponding non-encoded circuit when the error rate is below the threshold. In contrast, when the error rate is above the threshold, no advantage is observed in the fault-tolerant implementation. The developed high-accuracy optical system may provide a reliable platform to investigate error propagation in more complex circuits with fault-tolerant gates.\\
\end{abstract}

\maketitle

Error is inevitable in practical quantum computing during encoding, operation, and decoding processes. Although quantum error has been experimentally investigated \cite{cory1998,knill2001,wineland2004,philipp2011,reed2012,nigg2014,waldherr2014,bell2014,pan2019} in different physical systems, the experimental demonstration of a complete fault-tolerant computation \cite{shor1995,steane1996,shor1996pra,beth1997} remains still a great challenge \cite{gott2016} by the necessity of several high-quality qubits operating through high-accuracy quantum gates. In a full fault-tolerant implementation, quantum information processing, particularly including a set of universal quantum computation gates, is additionally protected against errors. The correct probability of the output from encoded fault-tolerant quantum circuit is higher than that from the corresponding non-encoded circuit if the error rate of underlying hardware is below a threshold \cite{knill1996,ben2008,got2000,raussendorf2007,fukui2018}. Therefore, fault-tolerant encoding may be essential to achieve future large-scale quantum computation \cite{got2010,muller2017}.

Practical quantum advantage may be achievable even under the presence of errors by applying the noisy intermediate-scale quantum (NISQ) technology \cite{preskill2018}. Although a complete fault-tolerant computation with practical application is still beyond the reach of NISQ technology, fault-tolerant quantum circuits could be demonstrated in a small system \cite{gott2016,rosenblum2018,puri2019} to show the effectiveness of noise mitigation in logical qubits and even implement some practical applications with NISQ technology \cite{bravyi2020}.

In Ref. \cite{gott2016}, a special fault-tolerant protocol is proposed for a small system consisting of five qubits, in which one is regarded as an ancillary qubit and the others are used to encode logical qubits. In this protocol, encoding, decoding, and some gates, such as single-qubit Pauli operators $\sigma^x$ and $\sigma^z$, and the two-qubit controlled-not (CNOT) operator (these Clifford operators are not universal), can be implemented fault-tolerantly in logical space with the help of post-selection. It is shown that errors in this protocol cannot be corrected. Following this protocol, fault-tolerant error detection of encoding has been demonstrated in trapped ions \cite{linke2017}, superconducting qubits \cite{takita2017}, and the IBM 5Q chip \cite{vuillot2018}. However, the key aspect of the quantum circuit implementing with fault-tolerant operations, that is, the existence of the threshold of error rate, has not been explicitly demonstrated.

In this work, we experimentally demonstrate the threshold of error rate for quantum circuits formed with fault-tolerant gates encoded in an all optical setup. Based on the encoding method in \cite{gott2016}, we encode two logical qubits using four qubits which are implemented with the optical path information \cite{jsxu2016,jsxu2018}. The polarization of photons is regarded as an ancillary qubit. Besides the encoding process, we implement a single-qubit Hadamard gate and a two-qubit CNOT gate in the logical state space to form a complete quantum circuit and detect the output to experimentally determine the fault-tolerant threshold. We consider the bit-flip error on each physical qubit during the whole process. Our results demonstrate that when the error rate remains below the threshold, the probability to obtain correct output results of logical qubits in the circuit with fault-tolerant gates is higher than that of the corresponding circuit without encoding logical qubit. On the other side, if the error rate is above the threshold, no benefit is obtained from the fault-tolerant implementation.

\begin{figure*}[htb!]
  \centering
  \includegraphics[width=0.8\linewidth]{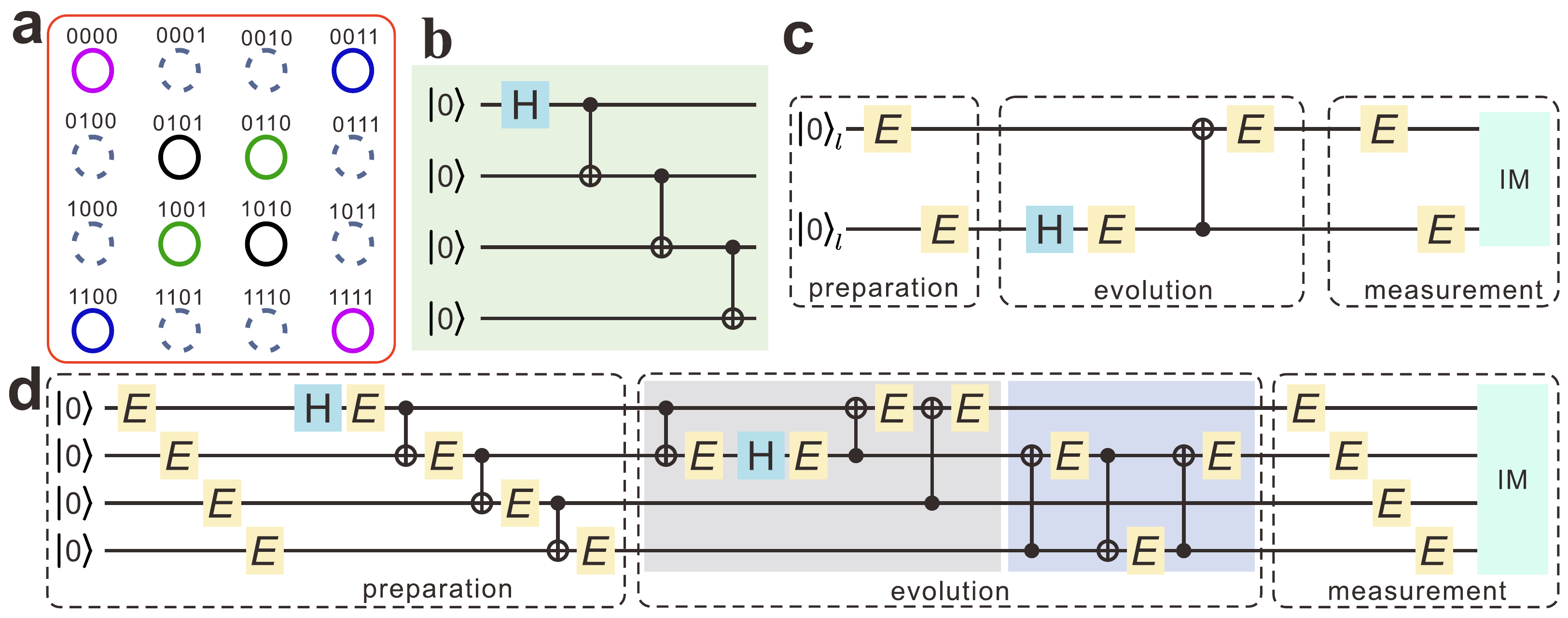}\\
  \caption{Quantum circuits with fault-tolerant gates under the presence of errors. {\bf a}. Spatial modes encoding basis states of four physical qubits. The states with odd number of 1s are indicated with dashed lines. Two modes with solid lines and the same color are encoded to be the corresponding logical basis. {\bf b}. The circuit of preparing logical qubit state $\ket{00}_l$ from the initial state $\ket{0000}$. {\bf c}. Non-encoded circuit including preparation, evolution and measurement starting from logical state $\ket{00}_l$. Hadamard gate $H_2$ applied to the second qubit followed by CNOT gate $CNOT_{21}$. Error gates $E$ are imported throughout the circuit. IM denotes ideal measurement. {\bf d}. The complete noisy physical circuit implementing logical operations with Hadamard gate $H_2$ and CNOT gate $CNOT_{21}$. In CNOT operation, error gate $E$ affects only the target qubit.}\label{fig:theory}
\end{figure*}

According to the fault-tolerant protocol in \cite{gott2016,linke2017}, two logical qubits are encoded with four physical qubits as follows:
\begin{align}\label{eq:logic-encode}
   &\ket{00}_l=(|0000\rangle + |1111\rangle)/\sqrt{2}, \notag \\
   &\ket{01}_l=(|0011\rangle + |1100\rangle)/\sqrt{2}, \notag \\
   &\ket{10}_l=(|0101\rangle + |1010\rangle)/\sqrt{2}, \notag \\
   &\ket{11}_l=(|0110\rangle + |1001\rangle)/\sqrt{2},
\end{align}
where $\{\ket{00}_l,\ \ket{01}_l,\ \ket{10}_l,\ \ket{11}_l\}$ represent the logical basis, and $\{\ket{0000}, \ket{0011}, \ket{0101}, \ket{0110}, \ket{1001}, \ket{1010}$, $\ket{1100},\ \ket{1111}\}$ represent the basis of four physical qubits (the encoded space only involves even number of $|1\rangle$ in physical qubits). The physical four-qubit basis is implemented with $2^4=16$ spatial modes. The correspondence between the spatial modes and the basis is illustrated in Fig. \ref{fig:theory}{\bf a} (the kets symbols are omitted brevity). Modes (dashed lines) including odd number of 1s are outside the encoded space. Each of the two modes (solid lines) with the same color represents each basis element of the physical qubits. By coherently moving spatial modes, single- and two-qubit gates can be conveniently realized (see the experimental part for more details). Logical state $\ket{00}_l=(\ket{0000}+\ket{1111})/\sqrt{2}$ can be fault-tolerantly prepared with post-selection following the circuit presented in Fig. \ref{fig:theory}{\bf b} starting from initial physical state $\ket{0000}$. In this protocol, a set of gates, such as $\sigma^x$, Hadamard and CNOT gates, operated on logical qubits can be implemented fault-tolerantly. As a result, a circuit, only formed by these fault-tolerant gates, is implemented fault-tolerantly and there exists a threshold of the error rate. Our main task is to experimentally demonstrate the existence of the threshold in the fault-tolerant circuit.

A complete fault-tolerant circuit includes preparation, evolution (containing a set of gates), and measurement, and an error may occur at any stage of the circuit. For simplicity, we describe an operator with error on qubits by its ideal quantum operation followed by an error gate $E$ (assumed to be the same for all the operations). The noisy measurement can be decomposed into error operation $E$ and the ideal measurement (IM). As illustrated in Fig. \ref{fig:theory}{\bf c} (non-encoded circuit), the evolution stage consists of a Hadamard operation $H_2$ on the second logical qubit followed by a two-qubit CNOT operation $CNOT_{21}$ in which the first (target) qubit is controlled by the second qubit. Fig. \ref{fig:theory}{\bf d} shows the corresponding encoded circuit implementing logical operations shown in Fig. \ref{fig:theory}{\bf c}.

The errors in preparation, evolution, and measurement are reflected in error gate $E$ in both non-encoded and encoded circuits, in which the error gate is assumed to be $E=\sigma^x$ with error rate $\epsilon=1-p$ (and $p$ being the success probability), thus establishing a coherent error. Experimentally controlling and identifying error rate $\epsilon$ in each gate is the key process in our study. The probability of correct output for each circuit can be defined as $F_p=\mathrm{Tr}[\rho_{ideal}\cdot\rho_{out}]$ (a function of $p$), where $\rho_{ideal}$ and $\rho_{out}$ are the ideal and experimental output states, respectively. To demonstrate the fault tolerance of circuit, we should confirm that the probability of correct output, $F_p$, in fault-tolerant circuit (Fig. \ref{fig:theory}{\bf d}) is higher than that in the corresponding non-encoded circuit (Fig. \ref{fig:theory}{\bf c}), $f_p$, obtained using the same hardware when success probability $p$ is above a threshold. The detailed calculations of $F_p$ and $f_p$ are presented in Ref. \cite{sm}.

\begin{figure*}[htbp]
  \centering
  \includegraphics[width=0.8\textwidth]{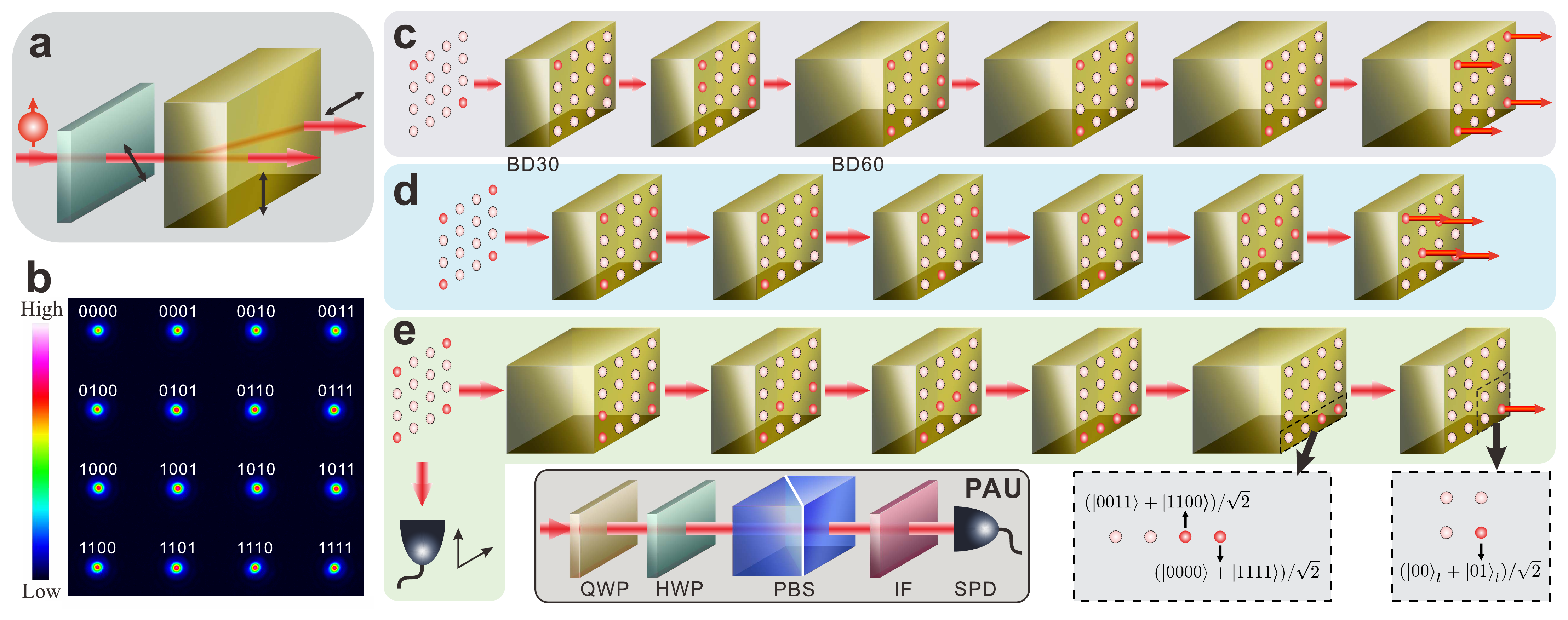}\\
  \caption{Experimental setup for verification of fault-tolerant threshold in quantum circuits.
  {\bf a}. Role of a beam displacer (BD). The polarization of photons prepared by a half-wave plate (HWP) is separated into two parallel spatial modes with horizontal and vertical polarizations.
  {\bf b}. Experimental image of optical spatial modes generated by exploiting a group of several BDs with different lengths (BD30 and BD60) and HWPs.
  The spatial mode evolutions of circuits implementing logical operations $H_2$ and $CNOT_{21}$ are shown in panels {\bf c} and {\bf d}, respectively.
  {\bf e}. Evolution of spatial modes for measurement at output state
  $(\ket{00}_l+\ket{01}_l)/\sqrt{2}$ after applying $H_2$. The explicit expression of spatial modes at the last two steps are shown in dashed panels. The coherent information in spatial modes is reflected in the polarization information, which is analyzed by a polarization analysis unit (PAU) consisting of a quarter-wave plate (QWP), a HWP, and a polarization beam splitter (PBS).
  }\label{fig:setup}
\end{figure*}

{\bf Experimental setup and results.}
The optical setup for this study is shown in Fig. \ref{fig:setup}. The input polarization photons are set by using half-wave plates (HWPs), and the spatial mode information is encoded by using different types of calcite beam displacers (BDs) that separate a beam into two parallel beams with orthogonal polarizations \cite{jsxu2016,jsxu2018}. The BDs are designed to separate the horizontal and vertical polarization components horizontally parallel, as shown in Fig. \ref{fig:setup}{\bf a}. Rotating the BD by $90^\circ$, it would lead to separate components vertically parallel. We use two types of BDs, namely, BD30 and BD60, which separate two beams by $3.0 mm$ and $6.0 mm$, respectively, to prepare $16$ spatial modes. The corresponding intensity image is shown in Fig. \ref{fig:setup}{\bf b}. Using photon polarization as the ancillary qubit and adjusting HWPs' angles,  amplitudes between different spatial modes change accordingly. The initial state can be prepared with a very high fidelity.

The rich encoding paths in the experimental setup enable different circuits for physical qubits to realize the same operations on logical qubits. However, only some configurations are fault-tolerant \cite{sm}. Circuits are constructed to fault-tolerantly implement Hadamard gate $H_2$ and CNOT gate $CNOT_{21}$ shown in the evolution part in Fig. \ref{fig:theory} {\bf d} on the two logical qubits (details on the evolution of spatial modes are shown in Fig. \ref{fig:setup}{\bf c} and {\bf d}), respectively. To implement a different circuit, we just need to rotate BDs and HWPs. And for some cases, we need simply to add more sets of BDs and HWPs.

\begin{figure*}[htbp]
  \centering
  \includegraphics[width=0.8\textwidth]{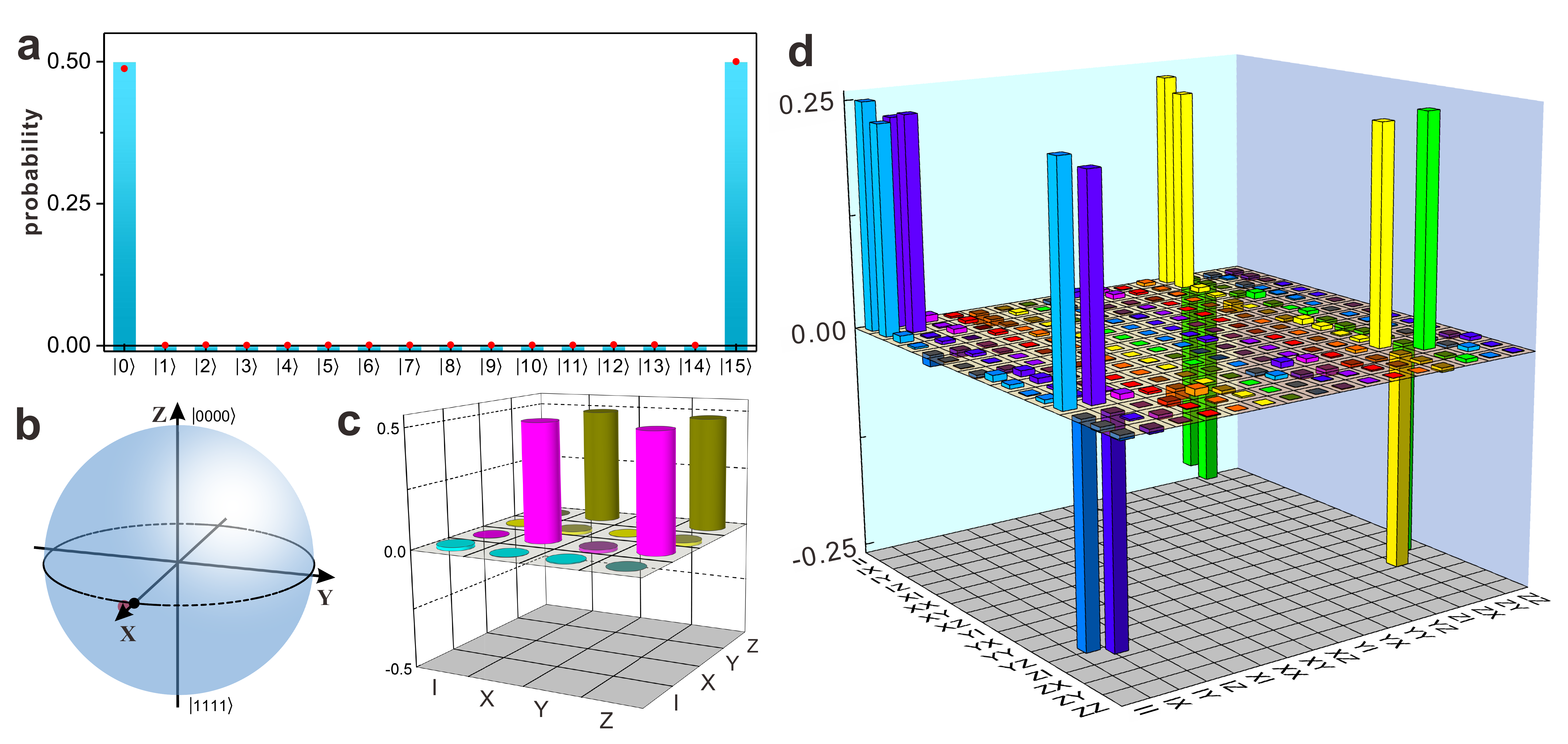}\\
  \caption{Experimental results of state preparation and quantum process tomography. {\bf a}. Detected probabilities of the complete basis for prepared logical state $\ket{00}_l$. The basis is represented with integers (e.g., $\ket{5}=\ket{0101}$). Histograms and red points correspond to the theoretical and experimental results, respectively. {\bf b}. Logical state
  $\ket{00}_l=(\ket{0000}+\ket{1111})/\sqrt{2}$
  represented in the Bloch sphere on basis $\{\ket{0000},\ \ket{1111}\}$. Black and pink points represent the theoretical prediction and experimental result, respectively. {\bf c}. Real part of density matrix of Hadamard operation. {\bf d}. Real part of density matrix of CNOT operation.
  }\label{fig:result1}
\end{figure*}

In the experiment, error gate $E=\sigma^x$, which flips the photon polarization, is realized by fine adjustment of HWP angular deviation and verified through measurements. To determine error rate $\epsilon=1-p$ experimentally, we first obtain probability distributions of spatial modes, which are detected directly by a two-dimensional movable single-photon detector (SPD) equipped with an interference filter (IF) (not shown) whose full width at half maximum is 3 nm as shown in Fig. \ref{fig:setup}{\bf e}. Success probability $p$ is then estimated by comparing the experimental probability distributions of all modes with the ideal prediction calculated through the error model \cite{sm}.

\begin{figure}[htb!]
  \centering
  \includegraphics[width=0.46\textwidth]{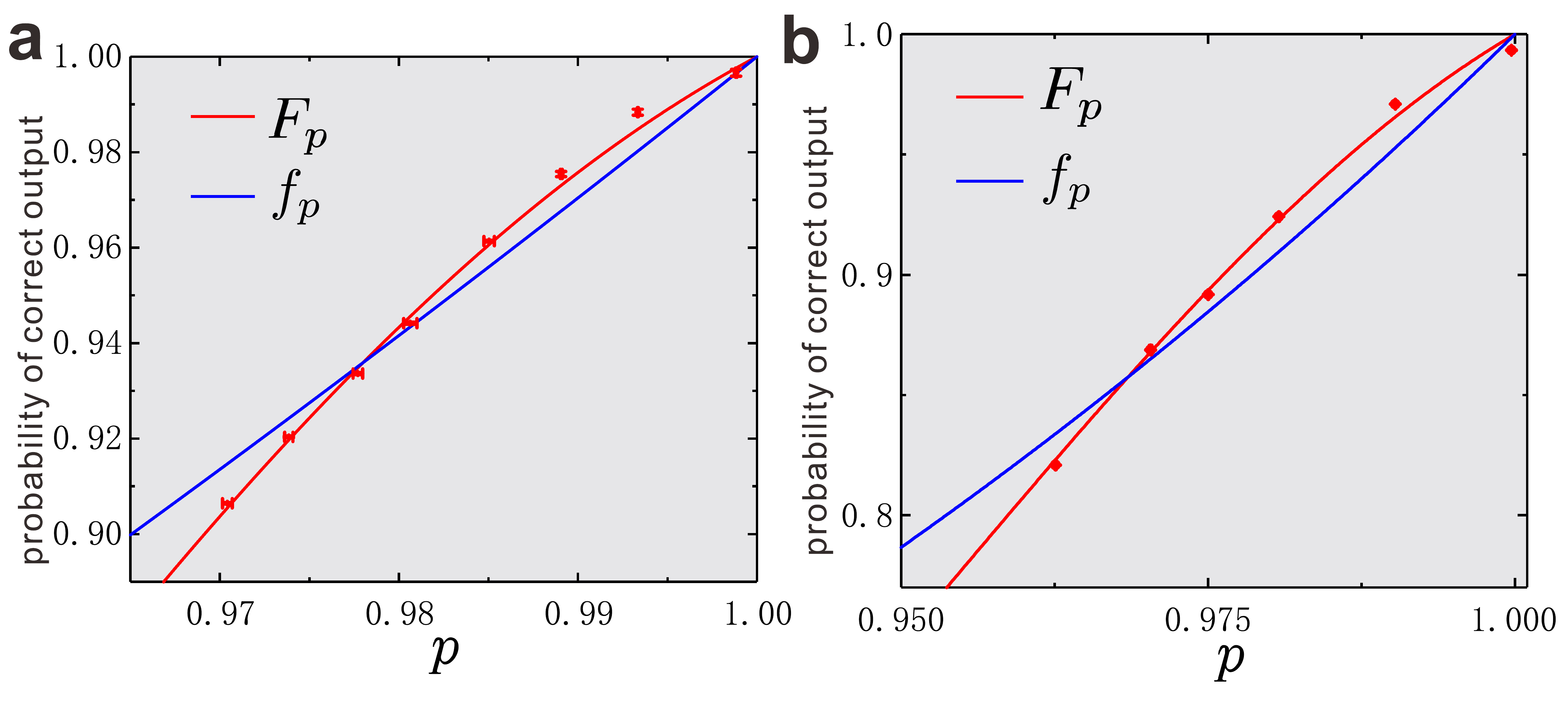}\\
  \caption{Experimental probabilities of correct output for different operations. Panels {\bf a} and {\bf b} show experimental probabilities of correct output, $F_p$, according to success probability $p=1-\epsilon$ for $H_2$ and $CNOT_{21} \cdot H_2$, respectively. Blue and red curves represent theoretical predictions of the non-encoded circuit ($f_p$) and fault-tolerant circuit ($F_p$), respectively. Red points with error bars indicate the experimental results. All error bars are estimated as standard deviations of photon counts assuming a Poisson distribution.
  }\label{fig:result2}
\end{figure}

The probability of correct output, $F_p$, can be obtained by projecting the output state onto ideal logical state basis. Concretely, we first obtain the total photon count $N_t$ which is the sum of eight modes in the encoding space with even number of 1s in the physical basis. The output spatial modes are then projected by the ideal logical state with photon count $N_i$ at the same period. The correct probability $F_p$ is thus given by $F_p=N_i/N_t$. Note that as both two counts are obtained in the same measured port, photon losses of optical elements are ignored. In experiments, the information encoded in the spatial modes is transferred to the polarization information. The measurement of the output state after $H_2$,  $|H_2\rangle_i=(\ket{00}_l+\ket{01}_l)/\sqrt{2}$, is illustrated in Fig. \ref{fig:setup}{\bf e}. Four spatial modes are stepwise combined with the combination of modes $\ket{0000}$ and $\ket{1111}$ representing logical state $\ket{00}_l$ and that of modes $\ket{0011}$ and $\ket{1100}$ representing $\ket{01}_l$. The polarizations of the two new modes, $\ket{00}_l$ and $\ket{01}_l$, are set to be horizontal and vertical, respectively. They are finally combined into one mode with the information encoded in the polarization degree of freedom. The similar method is used to measure the output state after $CNOT_{21}\cdot H_2$.

The polarization projective measurement is implemented by a polarization analysis unit (PAU), as shown in Fig. \ref{fig:setup}. During combination process, large optical path differences occur among several beams due to the unbalanced displacement. Lithium niobate (LiNbO$_3$) crystals and appropriate quartz plates are exploited to compensate the optical path difference \cite{sm}. The experimental projective measurement is realized with a successful probability of 99.2-99.8$\%$ with respect to the ideal projective measurement.

We first prepare logical basis $\ket{00}_l$ starting from initial state $\ket{0000}$ along the circuit shown in Fig. \ref{fig:theory}{\bf b}. Without importing error gates artificially, the detected probabilities of complete basis are shown in Fig. \ref{fig:result1}{\bf a}. The experimental results (red points) agree well with theoretical predictions (histograms). We deduce the inherent experimental error rate to be $\epsilon=0.0011\pm0.0001$ (i.e., $p=0.9989\pm0.0001$). The output state is reconstructed by quantum state tomography and depicted in the Bloch sphere in basis $\{\ket{0000},\ \ket{1111}\}$ with a probability of correct output of $99.50\pm0.01\%$ (Fig. \ref{fig:result1}{\bf b}).

We further demonstrate the high performance of Hadamard and CNOT gates on physical qubits in the proposed experimental setup. Quantum process tomography \cite{white2004} is performed, and the real parts of both gates are shown in Fig. \ref{fig:result1}{\bf c} and {\bf d}, with fidelities of operational matrices of $99.19\pm0.02\%$ and $97.59\pm0.01\%$, respectively. The corresponding imaginary parts of reconstructed density matrices are small, which are illustrated in Ref. \cite{sm}.

We then implement circuits with fault-tolerant quantum gates. The experimental reconstructed density matrices of output states of $H_2$ and $CNOT_{21}\cdot H_2$ are shown in Ref. \cite{sm}. The probability of correct output of Hadamard operation $H_2$ reaches $99.66\pm0.07\%$ with success probability $p=0.9988\pm0.0001$ (i.e., error rate $\epsilon=0.0012\pm0.0001$). The probability of correct output of $CNOT_{21}\cdot H_2$ reaches $99.33\pm0.02\%$ with success probability $p=0.9997\pm0.0001$ (i.e., error rate $\epsilon=0.0003\pm0.0001$).

The experimental results show that operations in the implemented platform are extremely accurate, allowing to observe the threshold effect in fault-tolerant protocol. Several probability distributions experimentally obtained from each spatial mode according to $p$ are shown in Ref. \cite{sm}. Results of $F_p$ according to $p$ are shown in Fig. \ref{fig:result2}{\bf a} for logical state $(\ket{00}_l+\ket{01}_l)/\sqrt{2}$ and Fig. \ref{fig:result2}{\bf b} for logical state $(\ket{00}_l+\ket{11}_l)/\sqrt{2}$. For the circuit implementing logical operation $H_2$, the threshold is $p=0.978$ in theory. This threshold is consistent with Fig. \ref{fig:result2}{\bf a}, in which the experimental probability of correct output, $F_p$, is larger than the prediction $f_p$ of non-encoded circuit for $p>0.978$. On the other hand, when $p<0.978$, we obtain $F_p<f_p$ experimentally. The experimental results of logical operation $CNOT_{21}\cdot H_2$ are shown in Fig. \ref{fig:result2}{\bf b}, in which the predicted threshold is $p=0.968$. The experimentally obtained $F_p$ is higher (lower) than corresponding $f_p$ for $p$ above (below) the threshold.

{\bf Conclusion.}
Using a concise fault-tolerant protocol, we experimentally demonstrate the threshold of a complete fault-tolerant circuit with a Hadamard gate and a CNOT gate on the logical qubits, besides preparing and measuring processes, with the bit-flip error in each operator. Generally, to verify a fault-tolerant protocol, the output error rate of any circuit, formed with fault-tolerant gates, should be lower than that of the corresponding non-encoded circuit when the error rate is below a threshold. To completely demonstrate a universal fault-tolerant quantum computation remains a long-standing challenging. The high-accuracy operations that can be achieved using optical systems establish a suitable platform to simulate the error propagation in fault-tolerant circuits, especially to investigate the behavior of coherent errors \cite{gott2016}. Also, based on this experimental platform, the similar protocol employing multi photons could be investigated under the same encoding framework \cite{sm}. Moreover, despite of limitation of the scale of optical system, this work facilitates the potential investigation of fault-tolerant protocols with the breakthroughs of large-scale experimental implementations of quantum technology based on ion-trap \cite{wineland2002}, superconductor \cite{barends2014,kelly2015,arute2019}, and ultra-cold atoms \cite{pan2020}.

\section{Acknowledge} This work was supported by the National Key Research and Development Program of China (Grants NO. 2016YFA0302700 and 2017YFA0304100), National Natural Science Foundation of China (Grant NO. 11874343, 11821404, 11774335, 61725504, 61805227, 61805228, 61975195, U19A2075), Anhui Initiative in Quantum Information Technologies (Grant NO.\ AHY060300 and AHY020100), Key Research Program of Frontier Science, CAS (Grant NO.\ QYZDYSSW-SLH003), Science Foundation of the CAS (NO. ZDRW-XH-2019-1), the Fundamental Research Funds for the Central Universities (Grant NO. WK2030380017, WK2030380015 and WK2470000026), the CAS Youth Innovation Promotion Association (No.\,2020447).

\section{Appendix}
\section{Analysis of circuits with error gates}
\subsection{The non-encoded protocol for the Hadamard operation on the second logical qubit}

First, we consider the non-encoded protocol in which the Hadamard operation is performed on a two-qubit state. The circuit is shown in Fig. 1{\bf c} in the main text. The initial state could be written $\rho_{in}=\ket{00}\bra{00}$. The Hadamard gate operation on the second qubit $H_2$ is written as
\begin{equation}\label{h2}
  H_2=\sigma^I\otimes H,
\end{equation}
with
$\sigma^I=\left(\begin{smallmatrix}
1 & \  0\\
0 & \  1\\
\end{smallmatrix}\right)$
being the identical operation and Hadamard gate
$H=1/\sqrt{2}\left(\begin{smallmatrix}
1 & \  1\\
1 & \  -1\\
\end{smallmatrix}\right)$.
For the ideal situation without error gates, the final state is
$\rho_{ideal}=H_2\cdot\rho_{in}\cdot H_2^\dag$.

Considering the error gate $E=\sigma^x$ illustrated in Fig. 1{\bf c} occurring in all the processes of preparation, evolution and measurement, the subsequent evolved state before operation $H_2$ is
\begin{equation}\label{eq:input-err}
\begin{split}
   &\rho_{in1}=p \rho_{in} + (1-p) E_1 \rho_{in} E_1^\dag, \\
   &\rho_{in2}=p \rho_{in1} + (1-p) E_2 \rho_{in1} E_2^\dag,
\end{split}
\end{equation}
where $1-p=\epsilon$ is the error rate of the error gate $E$ and $p$ represets the success probability.  $E_1=\sigma^x\otimes\sigma^I$ and $E_2=\sigma^I\otimes\sigma^x$
are the error gates implemented on the first and second qubits, respectively.

For the operation of $H_2$, the error gate acts on the second qubit. The evolution of state becomes
\begin{equation}\label{eq:h2-err}
\begin{split}
   &\rho_{ev1}=H_2 \rho_{in2} H_2^\dag, \\
   &\rho_{ev2}=p \rho_{ev1} + (1-p) E_2 \rho_{ev1} E_2^\dag.
\end{split}
\end{equation}

In the measurement process, both qubits are affected by the error gates. And the output state $\rho_{out}$ before the ideal measurement is
\begin{equation}\label{x1em}
\begin{split}
   &\rho_{m1}=p \rho_{ev2} + (1-p) E_1 \rho_{ev2} E_1^\dag, \\
   &\rho_{m2}=p \rho_{m1} + (1-p) E_2 \rho_{m1} E_2^\dag, \\
   &\rho_{out}=\rho_{m2}.
\end{split}
\end{equation}

By projecting output state $\rho_{out}$ on the ideal output state $\rho_{ideal}$, the correct probability $f_p=Tr[\rho_{out}.\rho_{ideal}]$ is theoretically calculated to $f_p=p-2p^2+2p^3$.

\subsection{Fault-tolerant circuits with error gates}
For the fault-tolerant circuit shown in Fig. 1{\bf d} in the main text. The correct probability $F_p$ is obtained in a similar way of the above non-encoded circuits by projecting the output state $\rho_{out}$ on the ideal measurement state $\rho_{ideal}$.

\begin{figure}[htbp]
  \centering
  \includegraphics[width=0.47\textwidth]{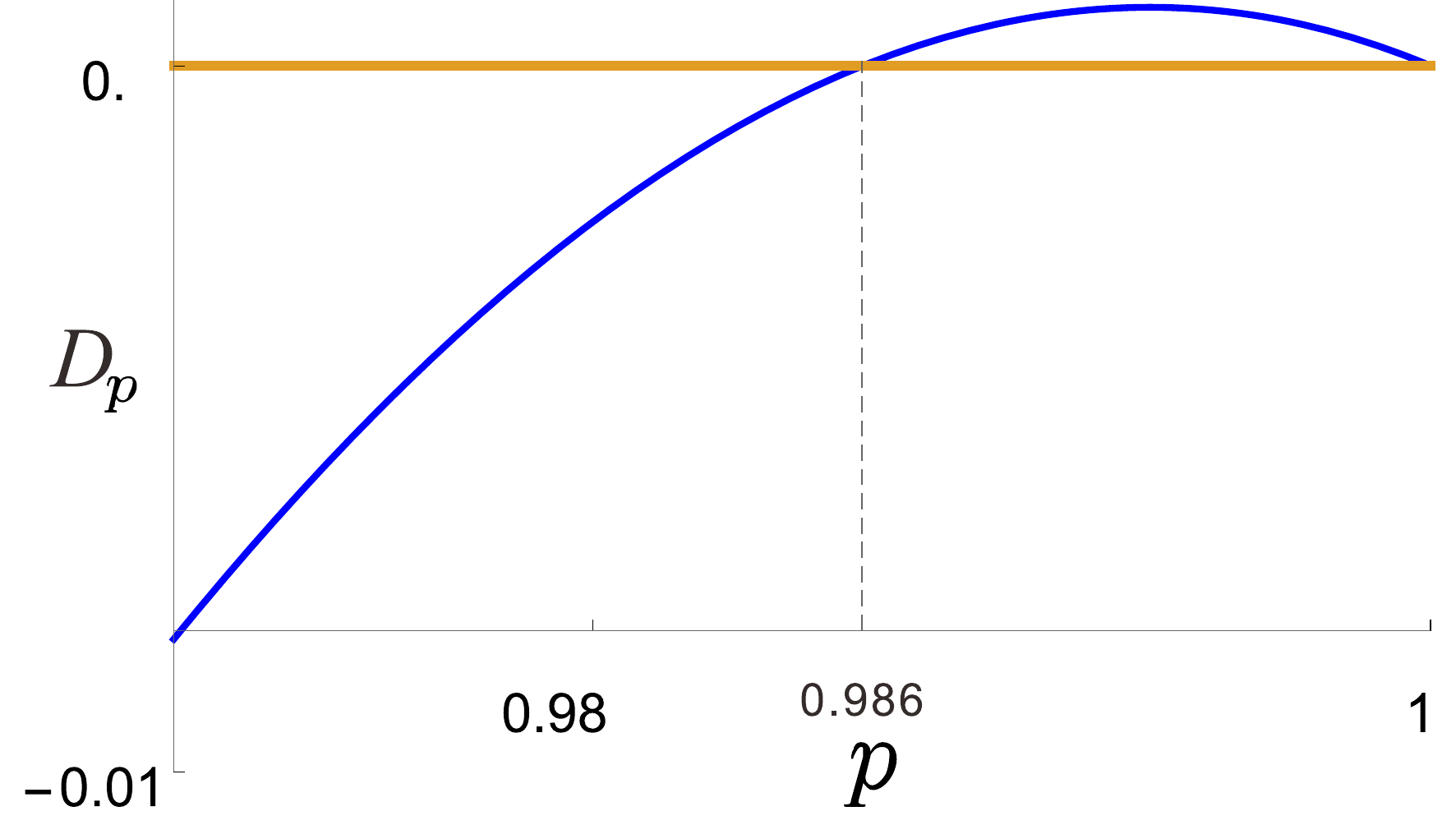}\\
  \caption{The difference $D_p=F_p-f_p$ between the correct probabilities of the fault-tolerant circuit and the non-encoded circuit for the preparation of logical state $\ket{00}_l$. The threshold appears at p=0.986.
  }\label{fig:preparation}
\end{figure}

For the preparation of logical state $\ket{00}_l$ whose circuit is shown in Fig. 1{\bf b} in the main text, following the error gates and measurement shown in Fig. 1{\bf d}, we can calculate the corresponding threshold about $p=0.986$. The results of $D_p=F_p-f_p$ are shown in Fig. \ref{fig:preparation}.

For the operation of Hadamard gate on the second logical qubit, compared with the theoretical prediction of non-encoded protocol, $F_p$ is larger than $f_p$ for $p>0.978$. The threshold error rate $\epsilon$ is about $0.022$. The comparison of $f_p$ and $F_p$ is shown in Fig. 4{\bf a} in the main text.

We further investigate the operation of $CNOT_{21}\cdot H_2$ in which $f_p=p-4p^2+12p^3-16p^4+8p^5$ for the non-encoded circuit. The comparison of $f_p$ and $F_p$ is shown in Fig. 4{\bf b} with the threshold being $0.968$.

During the evolution, the probability distribution of every mode can be calculated step by step.

\section{Analysis of different situations}

In the fault tolerance protocol, there is not just one encoded method to implement the circuits. Here, we introduce several other circuits to implement the logical operations.

\subsection{Different circuits}
\begin{figure*}[htbp]
  \centering
  \includegraphics[width=0.87\textwidth]{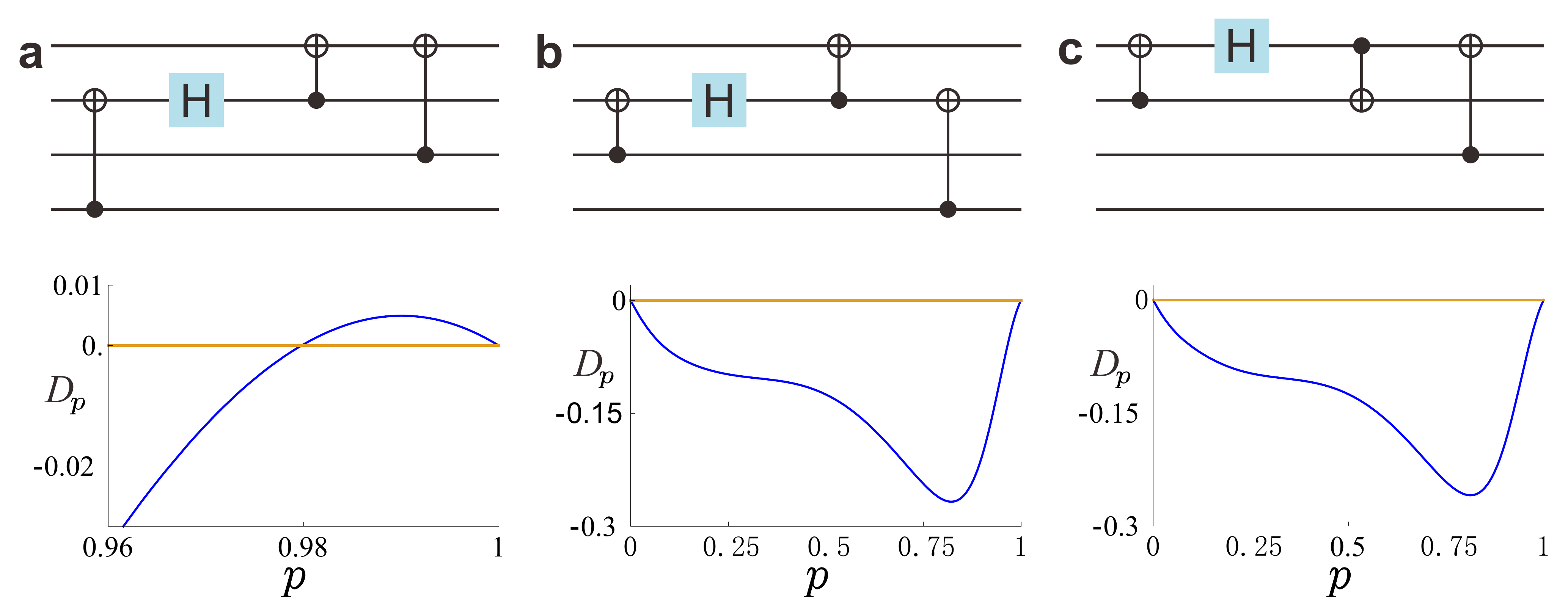}\\
  \caption{Different circuits to implement the logical operation $H_2$. The difference $D_p=F_p-f_p$ between the correct probabilities of the fault-tolerant circuit and the non-encoded circuit is shown below the corresponding circuit. The blue lines represent the theoretical results and the yellow lines represent the boundary of zero. The threshold in {\bf a} appears at p=0.98. While there are no thresholds in {\bf b} and {\bf c}.
  }\label{fig:s7}
\end{figure*}

Based on the preparation circuit introduced in the main text, other three circuits to implement $H_2$ are shown in Fig. \ref{fig:s7}{\bf a-c}. In these circuits, the operations on the four physical qubits are different from that shown in main text. By following the similar method introduced above to implement the error gate $E=\sigma^x$, theoretical predictions $F_p$ could be obtained. The corresponding differences between the correct probability of fault-tolerant and non-encoded circuits $D_p=F_p-f_p$ are shown under Fig. \ref{fig:s7}{\bf a-c}. For these circuits with error gates, different output states are generated and the corresponding thresholds sensibly change. For the circuits in Fig. \ref{fig:s7}{\bf b} and {\bf c}, $D_p$ is always less than 0 ($D_p=0$ for the trival cases with $p=0$ and 1), which means that there is no advantage for employing such encoded circuits.

\begin{figure}[htbp]
  \centering
  \includegraphics[width=0.47\textwidth]{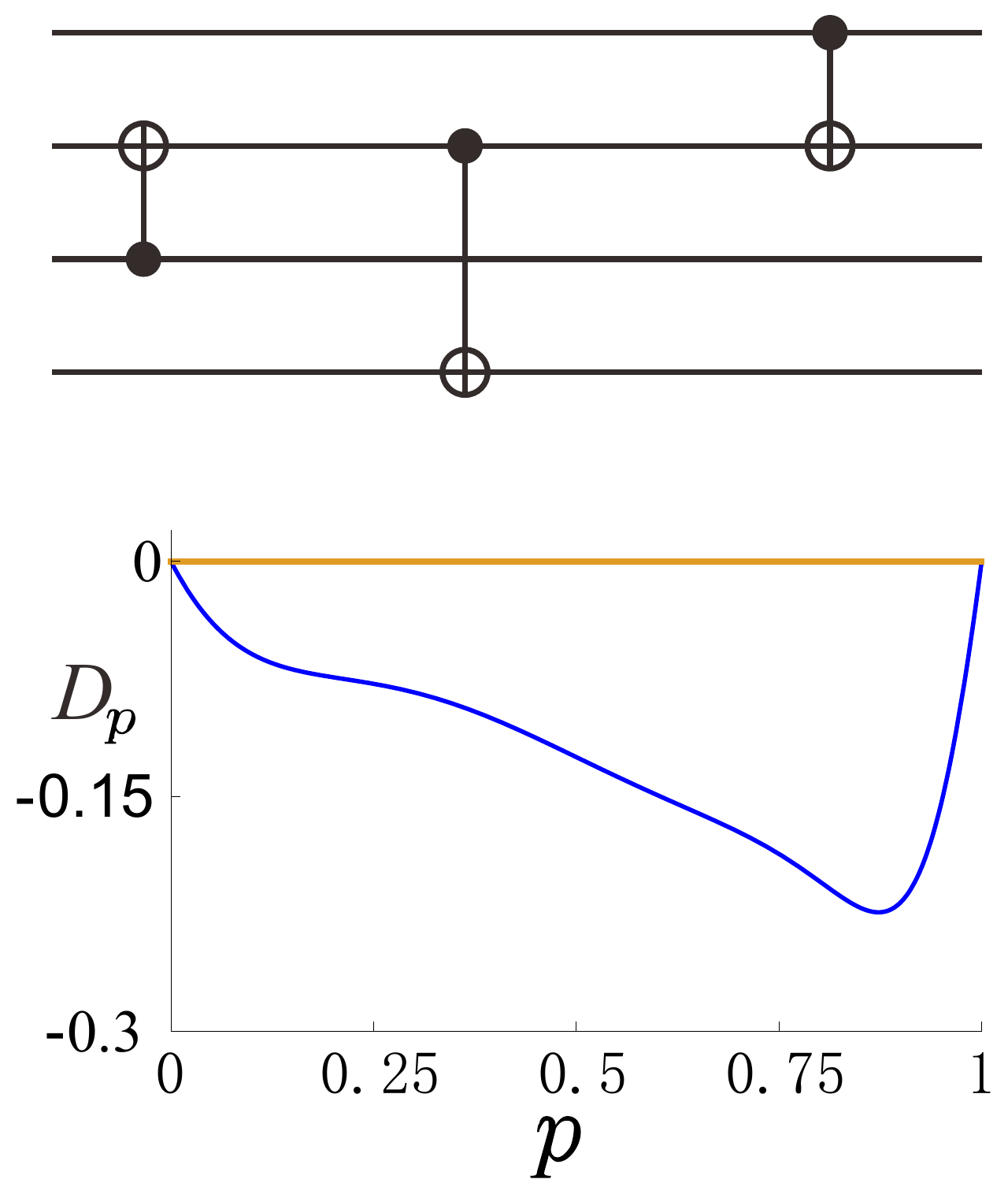}\\
  \caption{The circuit to implement logical operation $CNOT_{21}$. The difference $D_p=F_p-f_p$ is shown below the corresponding circuit. The blue line represent the theoretical result and the yellow line represent the boundary of zero.
  }\label{fig:s8}
\end{figure}

We further consider different circuits to implement the logical operation $CNOT_{21}\cdot H_2$ which is shown in Fig. \ref{fig:s8}. $D_p$ is always smaller than zero except for the trivial cases with $p=0$ and 1. As a result, there is no advantage for such encoded circuit.

\subsection{Different error gates}
\begin{figure*}[htbp]
  \centering
  \includegraphics[width=0.87\textwidth]{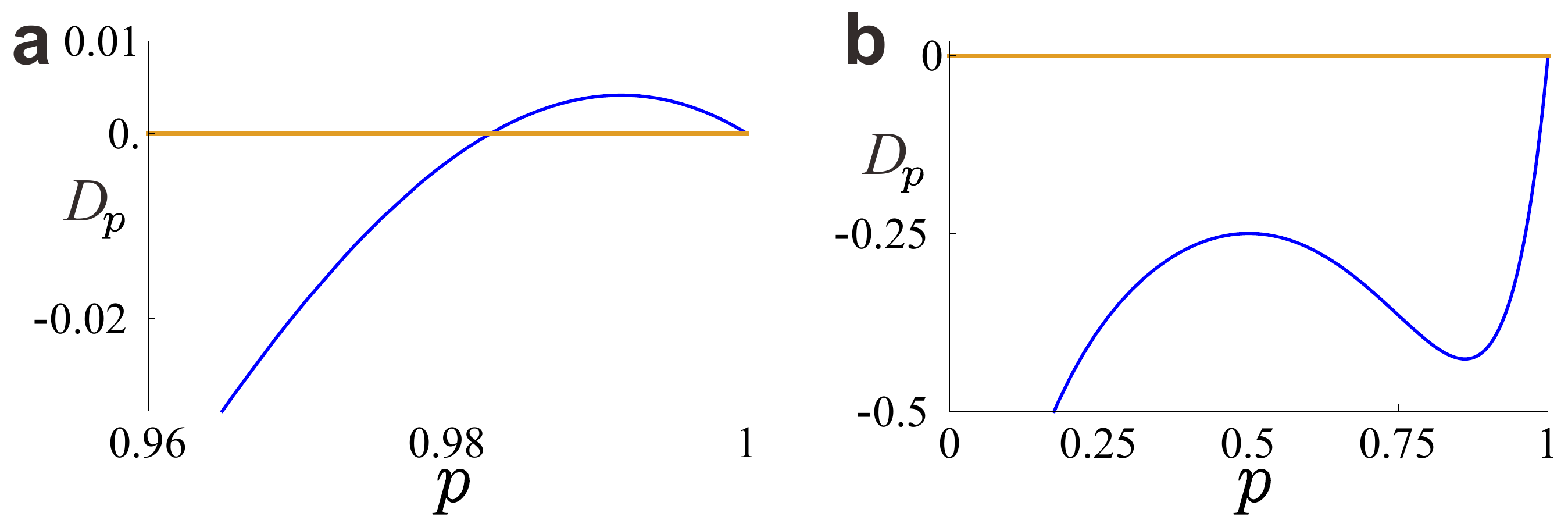}\\
  \caption{{\bf a}, {\bf b} show the theoretical results of $D_p=F_p-f_p$ for $E=\sigma^y$ and $E=\sigma^z$, respectively. The threshold in {\bf a} is about $p=0.983$.
  }\label{fig:s9}
\end{figure*}

Under the framework of encoding rules in Ref. \cite{gott2016}, we can investigate the gates, for both operation gates and error gates, which contribute the flip of qubit. For the phase gate which affects the phase between different kets, the number of 1s of the basis would not change, which leads all the physical states inside the encoding space. To clarify this point, we further consider two other types of error gate $\sigma^y$ and $\sigma^z$ to investigate their effect on the threshold of the fault-tolerant protocol. The used encoded circuit is same as that used in the main text for the Hadamard gate $H_2$. Fig. \ref{fig:s9}{\bf a} shows the theoretical results of $D_p=F_p-f_p$ with a threshold of $p=0.983$ for the error gate $E=\sigma^y$. Fig. \ref{fig:s9}{\bf b} shows the corresponding results when the error gate $E=\sigma^z$. Since the difference $D_p$ is smaller than zero for $p \in (0,1)$, there is no advantage for the encoded circuit compared with non-encoding circuit with $\sigma^z$ errors, which means the circuit is not fault tolerant.

\subsection{Different input state}
In our work, the input physical state starts from the generally initial one $\ket{0000}$. As mentioned in the main text, in our special encoding, only Clifford gates, such as CNOT, Hadamard and $X$ gates, can be implemented as the fault-tolerant manner. With these fault-tolerant gates, the logical input can be prepared as state $\ket{00}_l$. Obviously, we can implement the fault-tolerant gates on the initial state $\ket{00}_l$ and transform it to some new input states. For example, to get the input state $\ket{10}_l$, we just perform $X_1$ gate on the first logical qubit. We need to note that for each circuit formed with the fault-tolerant gates, there may exist a different error rate threshold. As a result, compared with the circuits, $CNOT_{21}\cdot H_2$ in our experiment, for the input state $\ket{10}_l$, the threshold will change since the $X_1$ gate could be treated as another logical circuit, which means the whole logical circuits $CNOT_{21}\cdot H_2 \cdot X_1$.

\section{More experimental details}

In the experimental setup, attenuated laser pulses, whose width is about 130 fs and repetition rate is about 76 MHz at the wavelength of 800 nm, with about $0.004$ photons averagely in every pulse are used as the photon source of input beam.

\subsection{Details of projective measurement}
In experiment, we project the final state to the logical basis by combining corresponding spatial modes and projecting to a proper polarization state. For example, by setting the spatial modes $\ket{0000}$ and $\ket{1111}$ to be horizontal ($\ket{h}$) and vertical ($\ket{v}$) polarizations, respectively, the correct probability of $\ket{00}_l=(\ket{0000}+\ket{1111})/\sqrt{2}$ can be obtained by combining the two modes together and projecting on the polarization state $(\ket{h}+\ket{v})/\sqrt{2}$ using the polarization analysis unit shown in the main text.For $(\ket{00}_l+\ket{01}_l)/\sqrt{2}$, we need to combine the modes $\ket{0000}$ and $\ket{1111}$ (the logical state $\ket{00}_l$), and the modes $\ket{0011}$ and $\ket{1100}$ (the logical state $\ket{01}_l$), respectively. If the phase information between $\ket{0000}$ and $\ket{1111}$ keeps stable, the state $\ket{00}_l$ which is originally in the polarization $(\ket{h}+\ket{v})/\sqrt{2}$ could be set to be $\ket{h}$, while $\ket{01}_l$ could be reset to be $\ket{v}$. Combining both two new modes to be $(\ket{h}+\ket{v})/\sqrt{2}$, we then project the state to $(\ket{00}_l+\ket{01}_l)/\sqrt{2}$.

\subsection{Compensation of the interferometer}
\begin{figure*}[htbp]
  \centering
  \includegraphics[width=0.67\textwidth]{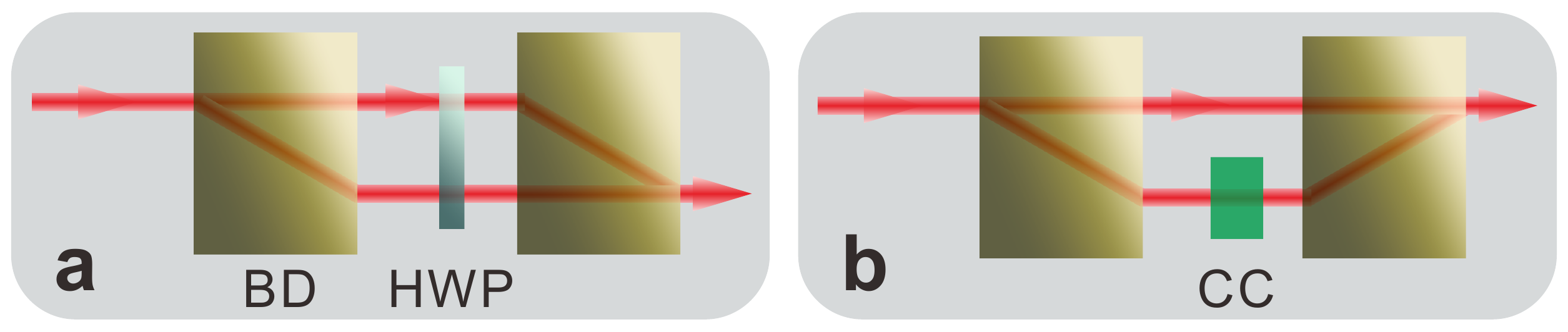}\\
  \caption{{\bf a}. Illustration of a balanced interferometer built up by two beam displacers (BDs) with a half wave plate (HWP) set at $45^\circ$. {\bf b}. An unbalanced interferometer with compensation crystal (CC) inserted.
  }\label{fig:s3}
\end{figure*}

In the combination of optical modes, beam displacers (BDs) are used to constitute the interferometer. As shown in Fig. \ref{fig:s3}, in a balanced interferometer (shown in Fig. \ref{fig:s3}{\bf a}) constructed by two BDs with a half wave plate (HWP) inserted at $45^\circ$, the two beams have the same optical lengths. Since these two beams are close to each other which suffers the same environmental noise, this kind of interferometer is inherently stable. While for the unbalanced interferometer shown in Fig. \ref{fig:s3}{\bf b}, a compensation crystal (CC) is placed on the path with a shorter optical length. In experiment, the optical path difference between two beams from a BD with a length of $28.3$mm is about $2.28$mm. A length of $4$mm Lithium niobate (LiNbO$_3$) crystal and several quartz plates are exploited as the CC inserted in the deflected beam to compensate the different optical length. In our work, the visibility of interferometer is very high to ensure that  experimental measurement is implemented with a successful probability of 99.2-99.8$\%$ compared with the ideal project measurement.

\subsection{Estimation of the value $\epsilon$}
\begin{figure*}[htbp]
  \centering
  \includegraphics[width=0.67\textwidth]{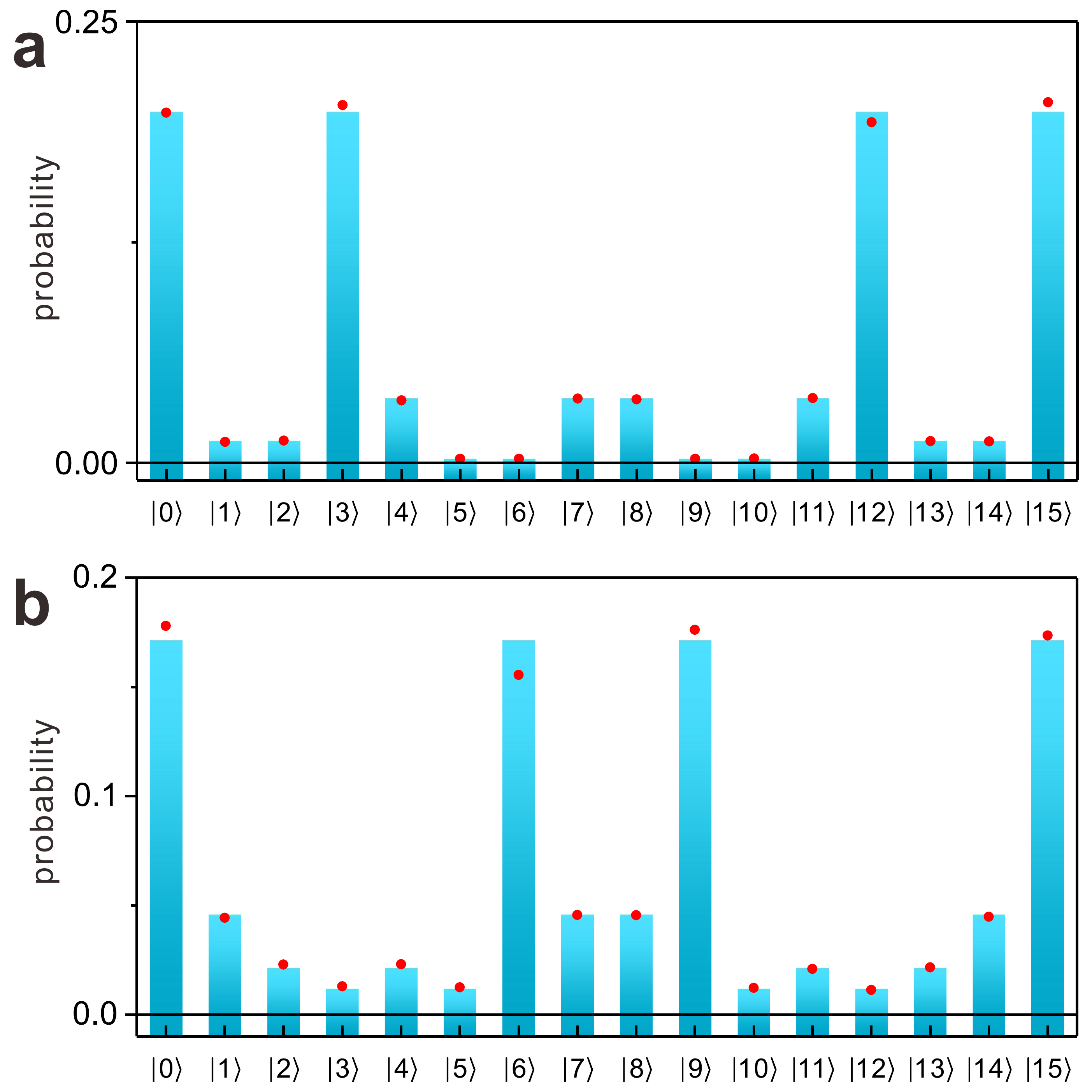}\\
  \caption{{\bf a}. Probabilities of 16 optical modes in the output of circuits with logical operation $H_2$ and the error rate is set to $\epsilon=0.03$. {\bf b}. Probabilities of 16 optical modes in the output of circuits with logical operation $CNOT_{21}\cdot H_2$ and the  error rate is set to $\epsilon=0.037$. The histograms and red points represent the theoretical and experimental results, respectively. The errorbars estimated to be standard deviations of the photon counts assumed to follow a Poisson distribution are too small to see here. The $x$ axis represents the basis of optical modes, which are denoted as the decimal form.
  }\label{fig:s4}
\end{figure*}

In experiment, the error rate $\epsilon$ is imported by adjusting the deviation of HWPs' angle. All intensities of the 16 optical spatial modes are detected with the two-dimensional movable detector, which is shown in the beginning of Fig. 2{\bf e}, to obtain the corresponding probability distributions. The experimental probability distributions of all modes are used to estimate the error rate $\epsilon$ by comparing with the ideal prediction calculated through the error model introduced in the above section I. Here we present some results of the probabilities in Fig. \ref{fig:s4}.

\subsection{More results of quantum process tomography}
\begin{figure*}[htbp]
  \centering
  \includegraphics[width=0.67\textwidth]{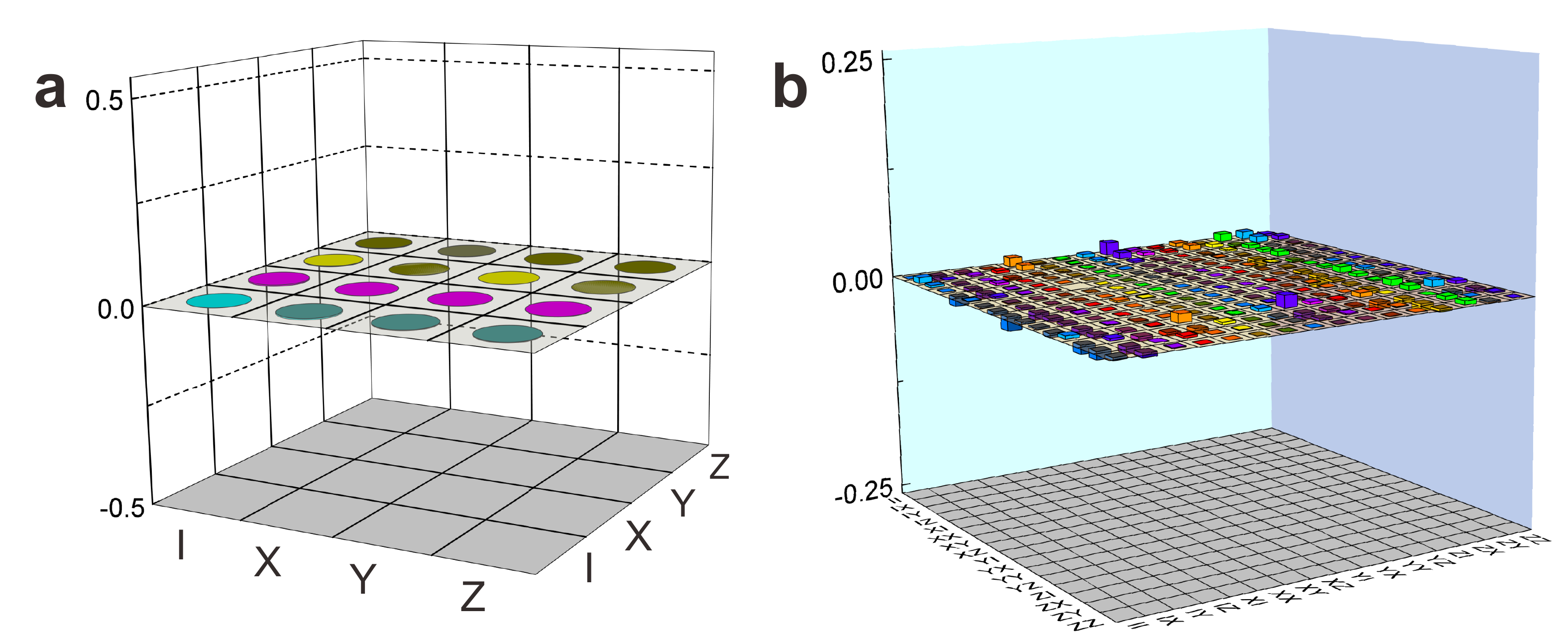}\\
  \caption{The experimental imaginary parts of density matrixes of Hadamard gate ({\bf a}) and CNOT gate ({\bf b}) on physical qubits based on the operational basis $\{I,\ X,\ Y,\ Z\}$.
  }\label{fig:s5}
\end{figure*}

The imaginary parts of density matrixes of Hadamard gate and CNOT gate on physical qubits based on the operational basis $\{I,\ X,\ Y,\ Z\}$ are shown in Fig. \ref{fig:s5}.
For the gate fidelities measured by quantum process tomography, the results are obtained by comparing the experimental-reconstructed density matrices of gates with the ideal matrices. It is quite different from the method of estimation of $\epsilon$. In the estimation of $\epsilon$, as mentioned above, the probability distributions of every mode, especially the probabilities of the modes owning odd number of 1s, play the core role. Since some modes will be projected outside the code-space, the coherent interactions between different modes are absent in the estimation of $\epsilon$. While for the gate fidelities, the coherence of different modes plays the key role. To obtain the predicted outputs, the measurements are more complex and are similar with the projective measurement on the logical state shown in Fig. 2{\bf e} in the main text. We need to compensate the optical path differences among the modes to constitute the interferometers which fulfill the corresponding coherence. And the imperfect interferometers in experiment affect greatly the quantum process tomography, which leads to lower the fidelities of gates.

\subsection{More results of output states}
\begin{figure*}[htbp]
  \centering
  \includegraphics[width=0.87\textwidth]{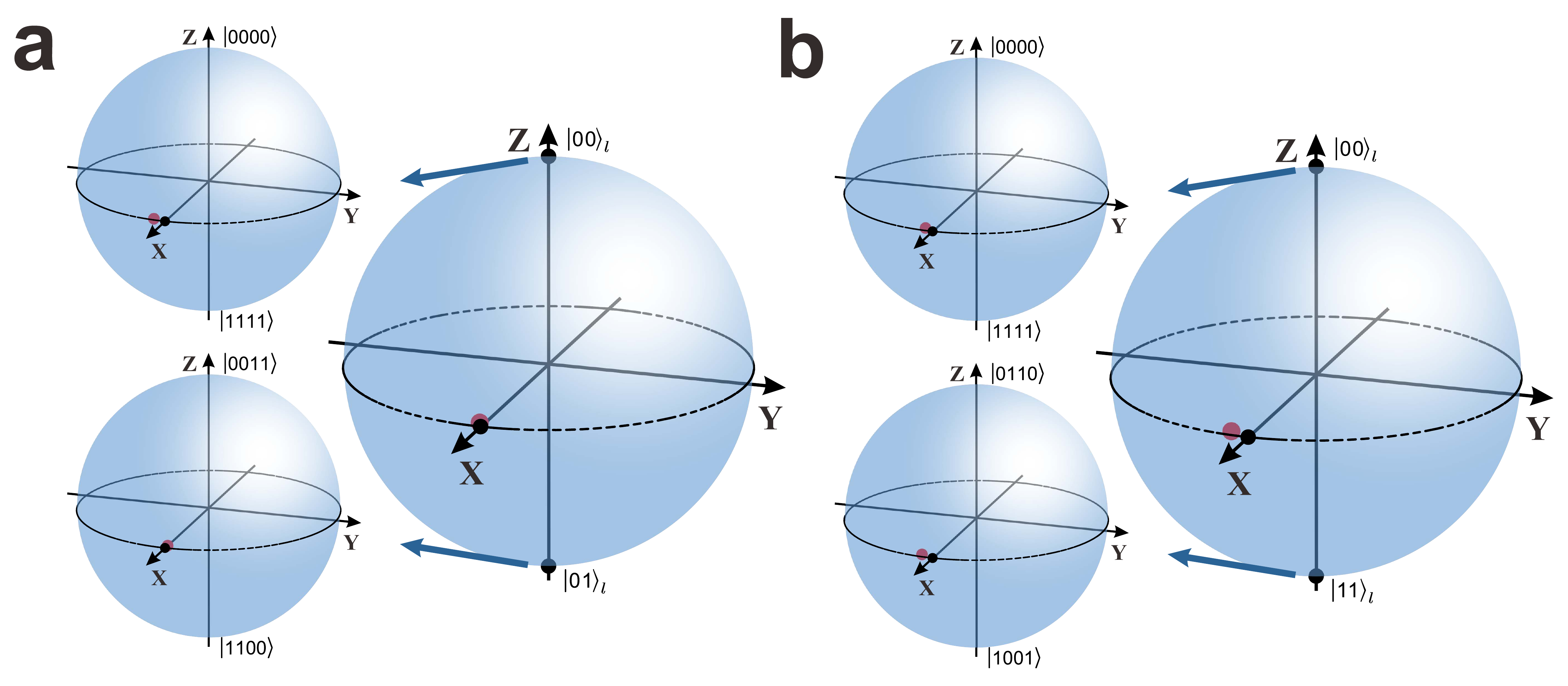}\\
  \caption{{\bf a} and {\bf b} show the experimental output states of logical operations $H_2$ and $CNOT_{21}\cdot H_2$, respectively, without importing the error rate. The state $(\ket{00}_l+\ket{01}_l)/\sqrt{2}$ is represented in the Bloch sphere based on $\{\ket{00}_l,\ \ket{01}_l\}$ in which $\ket{00}_l$ and $\ket{01}_l$ are shown in other two spheres on the relevant basis placed in the left. The output state $(\ket{00}_l+\ket{11}_l)/\sqrt{2}$ of logical operation $CNOT_{21}\cdot H_2$ along with $\ket{00}_l$ and $\ket{11}_l$ is shown in {\bf b}. The basis of $\ket{00}_l$ and $\ket{11}_l$ are denoted in the Bloch spheres with the basis of $\{\ket{0000}, \ \ket{1111}\}$
and
$\{\ket{0011}, \ \ket{1100}\}$, respectively. The black and red points correspond to the theoretical and experimental results, respectively.
  }\label{fig:s6}
\end{figure*}

Density matrix of the output logical state
$(\ket{00}_l+\ket{01}_l)/\sqrt{2}$
is reconstructed and shown in the Bloch sphere with the basis of $\{\ket{00}_l, \ \ket{01}_l\}$ in Fig. \ref{fig:s6}{\bf a}.
Here, logical states $\ket{00}_l=(\ket{0000}+\ket{1111})/\sqrt{2}$
and
$\ket{01}_l=(\ket{0011}+\ket{1100})/\sqrt{2}$ are reconstructed on the basis of
$\{\ket{0000}, \ \ket{1111}\}$
and
$\{\ket{0011}, \ \ket{1100}\}$,
respectively. The results are shown nearby the Bloch sphere of $\{\ket{00}_l, \ \ket{01}_l\}$ in Fig. \ref{fig:s6}{\bf a}. \ref{fig:s6}{\bf b} shows the results of the output state $(\ket{00}_l+\ket{11}_l)/\sqrt{2}$ on the basis of $\{\ket{00}_l, \ \ket{11}_l\}$ along with $\ket{00}_l=(\ket{0000}+\ket{1111})/\sqrt{2}$
and
$\ket{11}_l=(\ket{0110}+\ket{1001})/\sqrt{2}$ which are shown in the nearby Bloch sphere, respectively.

\section{The similar protocol using two entangled photons}

\begin{figure}[htbp]
  \centering
  \includegraphics[width=0.47\textwidth]{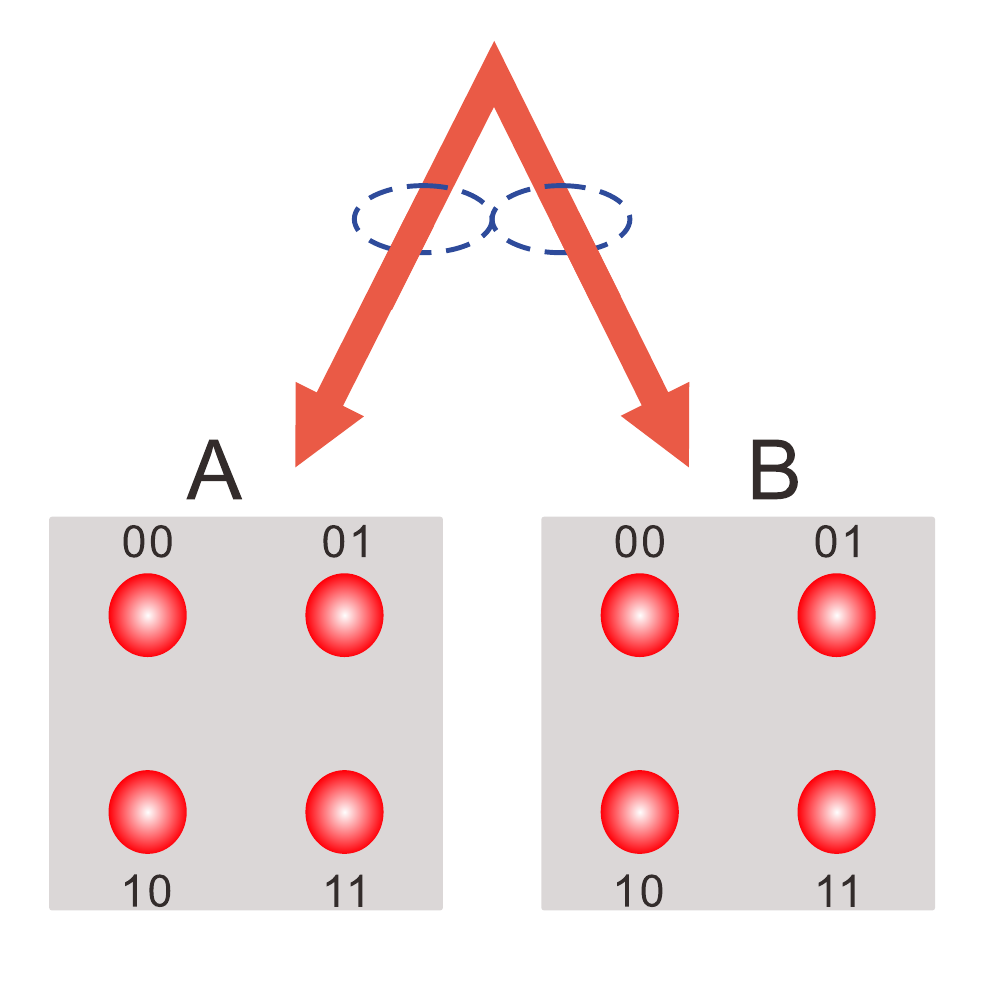}\\
  \caption{Illustration of the protocol using optical spatial modes of two entangled photons.
  }\label{fig:s2photon}
\end{figure}

Two polarized-entangled photons are sent to two sides, A and B, as shown in the Fig. \ref{fig:s2photon} here. The optical spatial modes are marked as $\ket{00}, \ket{01}, \ket{10}, \ket{11}$ on each side. The basis of four physical qubits is denoted as the combination of spatial modes on the sides of A and B. Here, $\ket{mnij}\equiv \ket{mn}_A \otimes \ket{ij}_B$ ($m, n \in \{0, 1\}_A$ and $i, j \in \{0, 1\}_B$). The photons on both sides share the maximally polarization entangled state $\ket{\Phi}_{AB}=1/\sqrt{2}(\ket{H_A H_B}+\ket{V_A V_B})$.

\begin{figure*}[htbp]
  \centering
  \includegraphics[width=0.88\textwidth]{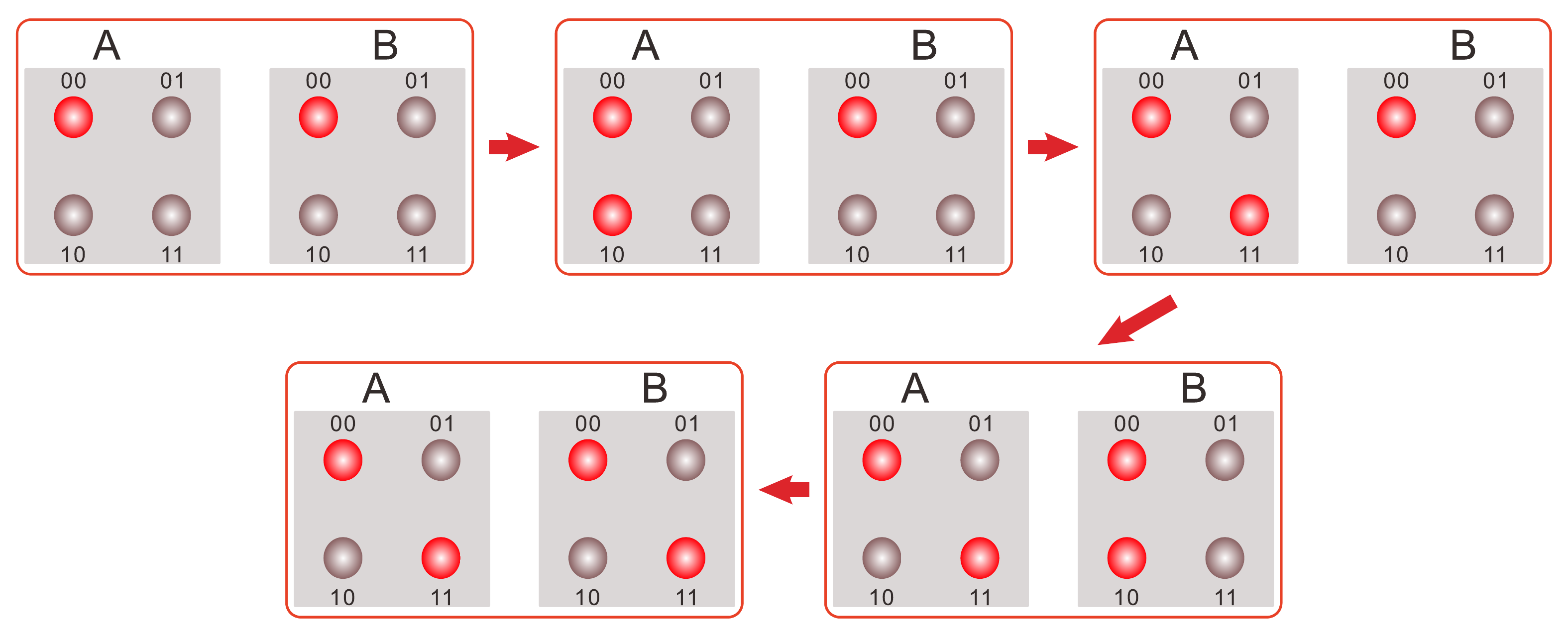}\\
  \caption{Evolution of the spatial modes on both sides in the preparation of logical state $\ket{00}_l$ starting from initial $\ket{0000}$.
  }\label{fig:s11}
\end{figure*}

The initial state of the optical spatial modes is in $\ket{0000}$ which equals $\ket{00}_A \otimes \ket{00}_B$. With the help of ancillary qubit - polarization, we could implement the operations shown in the main text. For example, the preparation of logical state $\ket{00}_l$ starting from initial $\ket{0000}$ is illustrated in Fig. \ref{fig:s11}.

For the initial state $\ket{0000}=\ket{00}_A \otimes \ket{00}_B$, the polarization of the photons in the modes of $\ket{00}_A$ and $\ket{00}_B$ are both along horizontal ($\ket{H}$) and vertical ($\ket{V}$). After the first vertical beam displacer (BD) on the side of A, the modes $\ket{00}_A$ splits into two modes with orthogonal polarizations, i.e., $\ket{00}_A$ in the polarization $\ket{H}$ and $\ket{10}_A$ in the polarization $\ket{V}$. This implements a Hadamard gate on the first physical qubit leading $\ket{0000}$ to the state $1/\sqrt{2}(\ket{00}_A \otimes \ket{00}_B + \ket{10}_A \otimes \ket{00}_B)$. With a horizontal BD in A's side, the state becomes $1/\sqrt{2}(\ket{00}_A \otimes \ket{00}_B + \ket{11}_A \otimes \ket{00}_B)$, which represents the result after the CNOT operation between the first and second physical qubits. For the CNOT operation between the second and third physical qubits, a vertical BD is added on B's side and the mode $\ket{00}_B$ splits into two modes with orthogonal polarizations, i.e., $\ket{00}_B$ in the polarization $\ket{H}$ and $\ket{10}_B$ in the polarization $\ket{V}$. Due to the entangled property, the state becomes $1/\sqrt{2}(\ket{00}_A \otimes \ket{00}_B + \ket{11}_A \otimes \ket{10}_B)$. Using another horizontal  BD on B' side, the final logical state $\ket{00}_l$ is prepared.

Similar operations can be implemented for the evolution and measurement processes, in which the results are obtained based on the coincidence counts between the detectors on both sides of A and B.

For this protocol, besides the coherent errors, the decoherence of two parties should be taken into consideration. If we still focus on the bit-flip error and ignore the decoherence, the theoretical predictions of the fault-tolerant threshold remain unchanged.

Similarly, if we use four entangled photons, we only use two spatial modes of every photon. However, it is totally predictable that the decoherence error, instead of the bit-flip error, plays a core role in the whole circuits.
\end{document}